\documentclass[12pt,preprint]{aastex}
\slugcomment{{\sc Accepted to ApJ:} August 10, 2010}
\usepackage{amsmath}
\usepackage{lscape}
\usepackage{rotating}
\usepackage{hyperref}

\shortauthors{Fuentes et al. 2010}
\shorttitle{HST Archival Search for TNOs}

\def\eq#1{\rm Equation (\ref{#1})}
\def\Eq#1{\rm Eq.~\ref{#1}}
\def\Fig#1{\rm Fig.~\ref{#1}}
\def\fig#1{\rm Figure~\ref{#1}}
\def\tab#1{\rm Table~\ref{#1}}
\def\km{~\rm km\ }

\def\au{~\rm AU\ }

\def\deg{~\rm deg}

\def\aph{~\rm ''~h^{-1}}

\def\pointing{{\it pointing}}
\def\pointings{{\it pointings}}
\def\eff{{\rm eff}}

\begin{document}
\bibliographystyle{apj}

\title{Trans-neptunian Objects with {\it Hubble Space Telescope}
  ACS/WFC\footnotemark[1]} \author{Cesar I.\,Fuentes\altaffilmark{2,3},
  Matthew J.\,Holman\altaffilmark{2}, David
  E.\,Trilling\altaffilmark{3}, Pavlos\,Protopapas\altaffilmark{2} }
\footnotetext[1]{Based on observations made with the NASA/ESA Hubble
  Space Telescope, obtained from the Data Archive at the Space
  Telescope Science Institute, which is operated by the Association of
  Universities for Research in Astronomy, Inc., under NASA contract
  NAS 5-26555. These observations are associated with program 11778.}
\altaffiltext{2}{Harvard-Smithsonian Center for Astrophysics, 60
  Garden Street, Cambridge, MA 02138, USA; cfuentes@cfa.harvard.edu}
\altaffiltext{3}{Department of Physics and Astronomy, Northern Arizona
  University, PO Box 6010, Flagstaff, AZ 86011 }

\begin{abstract}
We introduce a novel search technique that can identify
trans-neptunian objects in three to five exposures of a pointing
within a single {\it Hubble Space Telescope} orbit. The process is
fast enough to allow the discovery of candidates soon after the data
are available. This allows sufficient time to schedule follow up
observations with HST within a month. We report the discovery of 14
slow-moving objects found within 5$^\circ$ of the ecliptic in archival
data taken with the Wide Field Channel of the Advanced Camera for
Surveys. The luminosity function of these objects is consistent with
previous ground-based and space-based results.

We show evidence that the size distribution of both high and low
inclination populations is similar for objects smaller than 100\km, as
expected from collisional evolution models, while their size
distribution differ for brighter objects. We suggest the two
populations formed in different parts of the protoplanetary disk and
after being dynamically mixed have collisionally evolved together.
Among the objects discovered there is an equal mass binary with an
angular separation $\sim0\farcs53$.
\end{abstract}

\section{Introduction}\label{sec:int}
Trans-neptunian objects (TNOs) represent the leftovers of the same
planetesimals from which the planets in the solar system formed. These
offer a unique opportunity for testing theories of the growth and
collisional history of planetesimals and the dynamical evolution of
the giant planets \citep{Kenyon.2004, Morbidelli.2008}. The study of
the orbital distribution of TNOs has shown the existence of at least
two distinct dynamical populations \citep{Levison.2001, Brown.2001}
with different colors \citep{Doressoundiram.2008} and size
distributions \citep{Bernstein.2004, Fuentes.2008}.

Most of what is known about TNOs is based on follow-up studies of the
brightest objects \citep{Brown.2008}. The bias toward analysis of
brighter objects can be seen in challenging observations like
lightcurves and binarity fraction. This is even more apparent for
spectroscopic observations and albedo measurements, which are
available for only $\sim30$ objects \citep{Stansberry.2008,
  Brucker.2009}, among which the smallest is over 130\km in
diameter. This is explained by the relative faintness of outer solar
system bodies and the difficulty of tracking them after
discovery. Observations made several months and even years apart are
needed to secure accurate orbits. In general the fainter the object
the more demanding the observing conditions necessary to detect and
track it.

Despite the challenge, a great deal of effort has been dedicated to
searching for faint TNOs \citep{Chiang.1999, Gladman.2001, Allen.2002,
  Bernstein.2004, Petit.2006, Fraser.2008, Fuentes.2008, Fraser.2009,
  Fuentes.2009}. These surveys have concentrated near the ecliptic,
where the sky plane density of objects is largest. Elaborate
observational techniques have been developed to extend the sensitivity
of these surveys. Usually a compromise is reached between the sky
coverage and magnitude depth of these resource intensive
techniques. This results in ``pencil beam'' searches that concentrate
on a limited region of the sky. The results produced are statistically
calibrated and provide a precise assessment of the TNO sky plane
density. However these surveys typically obtain short arcs, yielding
imprecise information about TNO orbits.

These studies have extended our understanding of the TNO size
distribution to tens of km in diameter. In the deepest survey to date,
reaching a limit of $R\sim28.5$, \cite{Bernstein.2004} recognized a
break using HST data at $R\sim25$. At bright magnitudes the luminosity
function was consistent with the power law behavior surveys carried
out from the ground had measured. However those searches claimed the
luminosity function of bright objects could be extended up to a
magnitude $R\sim26$ \citep{Gladman.2001, Petit.2006}. The controversy
was settled when \cite{Fuentes.2008} corroborated the existence of the
break. That work had the advantage of being a single survey with the
sky coverage and magnitude depth to be sensitive to $R\sim 25.5$
objects and obtained a statistically significant result that did not
rely in the combination of fields observed under various
conditions. Deeper ground based searches have been able to narrow the
gap between ground and space based surveys by coadding data taken over
an entire night \citep{Fraser.2009, Fuentes.2009}.

\cite{Brown.2001} determined that the TNO inclination distribution was
well fit by the sum of a narrow and wide gaussian distribution.
\cite{Bernstein.2004} used the somewhat arbitrary value of $i=5^\circ$
to differentiate between hot and cold objects and recognized different
size distributions for both populations. They determined the size
distribution of hot objects had a shallower slope than that of cold
objects for objects larger and smaller than the break. Observationally
most large objects are hot and most small objects are cold.  However,
this result was based on a few objects smaller than the break,
especially on the three cold TNOs found by \cite{Bernstein.2004}.

A simple definition for hot and cold objects is useful for pencil beam
surveys where the constraint on the orbits is not precise. From the
ground not much more than a rate of motion on the sky is obtained from
a night's worth of observation. The large uncertainties associated
with the distance and inclination estimated under the assumption of a
circular orbit could eventually bias the analysis. A survey able to
find faint objects ($R\sim26$) and provide accurate constrains on the
distance and inclination could show if indeed there is a difference in
the size distribution of high and low inclination objects.

The most basic information that can be extracted from a set of TNO
discoveries is the luminosity function. If the albedo is assumed and
the distance to each object can be estimated, the size distribution is
obtained.  With further information about the trajectory of an object
we can estimate its inclination, which can be used as a proxy for
dynamical excitation. Ground based detections provide a very short arc
that gives us limited information about the distance if the degeneracy
between the object's velocity and parallactic motion cannot be
disentangled.

The HST has the advantage of not being affected by atmospheric seeing,
achieving very precise astrometric measurements. Also, its orbital
motion about the Earth adds extra parallax to the observations. For
Solar System bodies this helps in unraveling the contributions of the
Earth's parallax and the object's intrinsic motion, allowing precise
orbital estimates, even when not observing at opposition.

The objective of our investigation was to find faint TNOs with
acceptable orbital uncertainties to further constrain the size
distribution of the hot and cold populations. For this we defined a
limited, well characterized search for moving objects. Our search is
sensitive to $R\sim26$ and is able to constrain the distance and
inclination of the objects discovered. In Section \S\ref{sec:dat} we
present a summary of the data selection and acquisition. The
characterization of the search algorithm is done by sampling a control
population, described in \S\ref{sec:pop}. The detection pipeline is
described in detail in \S\ref{sec:mod}, and the detection efficiency
is explained in \S\ref{sec:eff}. Results and analysis of the data
appear in \S\ref{sec:ana}, where special emphasis is given to testing
the capabilities of HST in finding the correct orbital information. We
discuss the significance of our findings in \S\ref{sec:dis}. Our
conclusions appear in \S\ref{sec:con}.

\section{Data}\label{sec:dat}
Objects in the TNO realm ($\sim42\au$) exhibit parallactic motion of
$\sim3\aph$ when observed at opposition, mainly due to Earth's
translation. Depending on the resolution and data quality of the
observations TNOs are readily identified by this motion if two or more
images of the same field are taken with an adequate interval between
exposures. This parallactic motion implies that if the shutter is kept
open for a time longer than it takes a TNO to move beyond its PSF, the
image will trail. If observed at opposition, an image of a typical TNO
will take $\sim10$ min to traverse the PSF of a ground based image
(${\rm FWHM}\sim0\farcs5$) while only 1 min in an image taken with the
Wide Field Channel of the Advanced Camera for Surveys (ACS/WFC, ${\rm
  FWHM}\sim0\farcs05$).

We focused our search on data taken with ACS/WFC, the largest field of
view camera on HST ($202''\times202''$ or $0.003\deg^2$.)
\cite{Bernstein.2004} coadded tens of ACS/WFC exposures to reach a
sensitivity of $R\sim28.5$. However, of the three objects they
discovered two of them were detected in each individual image, and the
faintest ($R=27.8$) exhibited a lightcurve that made it visible in a
fraction of the exposures. The latest results for the TNO luminosity
function \citep{Fuentes.2009, Fraser.2009} indicate that the sky
density of TNOs brighter than $R=27$ on the ecliptic is 0.5 per
ACS/WFC field. The lack of the degrading effect of the atmosphere
compensates for the relatively small size of HST's 2.4m mirror.

The archive provides numerous data of different targets, with
different filters and science goals. The ACS/WFC data provided by the
Space Telescope Science Institute (STScI) is quite homogeneous in its
format which allows us to build software that can apply a standard
processing procedure for all data considered in this project. Most
exposure times are $\sim500$s in order to maximize the open shutter
time. In addition, it is customary that longer observations be divided
in a number of shorter exposures, allowing a median rejection of
cosmic rays. This, typically three or more exposures of a field are
obtained in sequence.

Access to the HST's electronic archive is provided by the Multimission
Archive at STScI (MAST) (\href{http://archive.stsci.edu}
{archive.stsci.edu}).

\subsection{Field Selection}
\begin{figure*}[Ht]
  \epsscale{1.0}
  \plotone{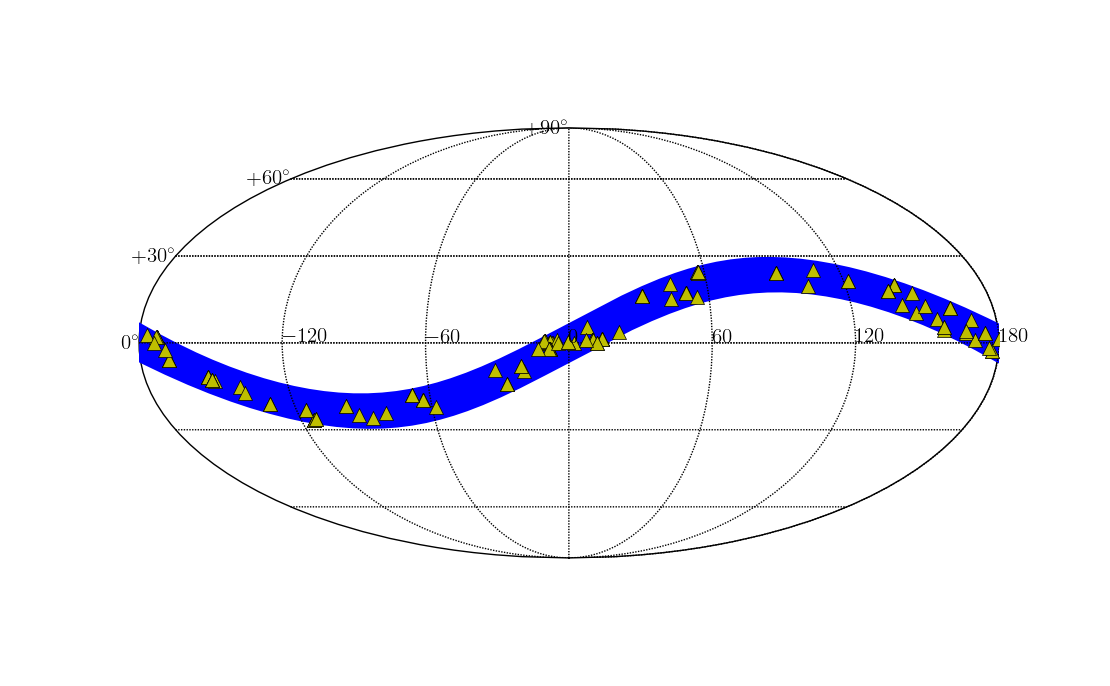} 
  \caption{\label{fig:fields} Map of the sky in J2000 coordinates. The
    $10^\circ$-wide   ecliptic   band   we   chose   to   select   our
    \pointings~from  is shown  in blue.  The location  of  all targets
    considered  is plotted  as  yellow triangles.  Many pointings  are
    superimposed on top of each other. }
\end{figure*}

We considered observations obtained within $5^\circ$ of the ecliptic,
where the sky density of TNOs is highest, as their orbits are
concentrated near the ecliptic \citep{Brown.2001}. \fig{fig:fields}
shows the distribution of fields we considered.

It is common for ground based surveys to prioritize fields located at
opposition. This maximizes the parallactic motion with respect to the
object's intrinsic velocity, allowing a reasonable $10-20\%$
uncertainty estimate on the distance if a circular orbit is
assumed. It also permits a clear distinction of nearer, main belt
asteroids from TNOs. Given the superior resolution of ACS/WFC data and
the extra parallax derived from the motion of HST itself it is not
necessary to observe at opposition to constrain the distance to a
moving object without having to rely on a circular orbit. For this
reason, we did not restrict out attention to a specific range of solar
elongation.

We consider images of the same field taken within the same HST orbit
as part of a \pointing. Only \pointings~that had a total open shutter
time of over 1,000 seconds with three or more images were
considered. These images are typically taken within half an HST
orbital period, $\sim48\min$. A total of 150 \pointings~were
recognized as satisfactory for this project. We specifically excluded
the many observations taken for the work by \cite{Bernstein.2004}, as
those were previously searched for TNOs.

STScI makes available data in any step of the reduction process. We
selected flat-fielded images that had not been undistorted or
combined. We used the distortion corrections and PSF models for
various filters provided by \cite{Anderson.2000}. The filters that we
considered for this work are summarized in \tab{tab:filters}.

\subsection{Astrometric Solution}
We search for Solar System objects that change position between images
over the time span of the \pointing. The best astrometry possible is
necessary to obtain a precise trajectory for TNOs. The astrometric
accuracy provided by the archive's calibration is only as good as the
astrometric precision of the HST guide star catalog provided for the
``astrometric reference''. However, the differential astrometry can be
as accurate as the HST's spatial resolution.

Instead of combining data taken in different \pointings, we took
advantage of the precise differential astrometry between images
obtained during the same orbit of the HST. ACS/WFC provides exquisite
resolution ($\sim50~$mas) and a stable and nearly constant PSF
\citep{Anderson.2000} across the field of view. There is, however a
significant large scale distortion, that needs to be accounted for
before detections in different images can be compared with each other.
In order to obtain a consistent astrometric solution for all the
images in a single \pointing~we used the distortion solution in the
software developed by \cite{Anderson.2000} that considers
filter-dependent distortions at the pixel level.

We defined the first image in the sequence as the ``astrometric
reference''. All images were searched for sources and their positions
transformed to the undistorted frame. All images in the \pointing~are
registered to the reference by matching common sources. All detections
are transformed to J2000 coordinates via the astrometric information
in the reference image provided by STScI. In this way there is a
unique transformation from a detection in any image to J2000
coordinates. This transformations from the image array to sky
coordinates is readily inverted to be used during the implanting of
the control population, to be discussed in the next section. For
static sources the uncertainty in the position was very close to the
50~mas advertised by the ACS/WFC's documentation.

\section{Control Population}\label{sec:pop}
Our moving object detection method requires the identification of an
object in at least three different images. There are various reasons
for an object in the field to go undetected. Being too faint is the
most common. Chance alignment with a background star or cosmic ray
(CR) will also reduce the chances of finding an object. Chip gaps and
bad pixels should be taken into account also when considering the
detection efficiency. We measure these and other unknown effects with
the use of a control population that covers the range of observational
characteristics the TNO population is expected to exhibit.

We implant our control population in the original flat-fielded images,
before any distortion correction is applied. Since these objects go
through the pipeline with the original data, anything that would
affect our ability to detect real faint objects will also affect our
ability to detect the objects that were implanted.

During the visual examination phase, to be discussed later, the
operator is presented with thousands of candidate moving objects. This
provides a constant stream of objects moving in TNO-like trajectories;
real objects are indistinguishable from the control population. For
the detection of new objects the most important characteristics to be
modeled are the brightness and rate of motion distribution for the
synthetic population.

\subsection{Apparent Motion}
The control population also provides a test for the reduction
pipeline, the recovery software, and the visual examination. The
analysis pipeline, including the human interaction, will be successful
if it discovers real objects that do not look special among the
control population. This means the control population should be
accurate and look like real TNOs. To avoid being biased toward finding
exactly what we expect, based on what we already know about TNOs, we
require a control population that spans all realistic properties (for
example, orbits, colors, lightcurves, and binarity.) However, for
simplicity, we considered only single (not binary TNOs), showing no
brightness variation in a $\sim40\min$ interval and with normal TNO
colors (See subsection~\ref{sec:phot}.) We only considered bound
orbits.

In order to have both an accurate and inclusive TNO control
population, we considered two different parameterizations. We first
used the Keplerian orbital elements of an object to produce
ephemerides. This allowed us to produce a distribution of orbital
parameters similar to that of TNOs. The second one was based on the
\cite{Bernstein.2000} elements which considers a cartesian grid
centered on the position of the observer at the time of
observation. These elements are closely related to, and therefore a
better measure of, the observational rates of motion. For this reason
this method is more inclusive. Equal number of objects, usually 200,
produced with each approach were implanted in each
\pointing. Ephemerides produced by these two methods using independent
pieces of software, a variation of \cite{Bernstein.2000}'s {\it
  Orbfit} and a custom made integrator, are used by the procedure that
inserts synthetic objects in the original images.

\subsection{Brightness Distribution}
For any given \pointing~we considered a uniform distribution in the
instrumental magnitude. The faint end of this distribution was
selected based on the reported instrumental zeropoint and exposure
times of the images. We selected the magnitude distribution so that it
would yield $\sim50$ detected objects per \pointing, enough objects to
sample the efficiency function of each field. The magnitude range
spans 2.5 magnitudes, and the faintest object was chosen to be half a
magnitude fainter than the faintest object that should appear as a
1-sigma detection in an individual image.

As objects will trail over the course of an integration, our software
computes the object's position at the beginning, middle and end of the
exposure based on its orbital parameters and the position of the HST
at the time of the exposure. We fit a 2nd-degree polynomial to this
motion and then subdivide that motion in 1-pixel increments. We then
divide the object's flux by the number of positions and insert a
normalized PSF model at each position for that particular filter
\citep{Anderson.2000}. Based on the position on the array we can also
correct the brightness of a source as another effect of the
geometrical distortion. Since the photometric uncertainty of the
objects that we are interested in are background-noise limited, no
additional noise was added to trailed PSFs.

\section{Detection of Moving Objects}\label{sec:mod}

The usual strategy for finding TNOs that are detectable in single
images relies in testing all correlations between detections across
images that are consistent with a TNO orbit. From the ground the image
quality is such that exposure times of some minutes can be used before
trailing is an issue. Then the we search for correlations between
point-like sources that move from image to image. For observations
taken over a single night of observation the algorithm takes the list
of detections and finds subsets that follow a straight line with a
constant rate.

The diffraction limited resolution with respect to the ground and the
apparent motion induced by HST orbiting the Earth imply that TNOs
detections will be trailed in typical exposure times ($500$ s). This
trailing spreads an object's flux over a larger number of pixels,
which for background limited observations significantly decreases the
likelihood of finding a faint moving object. For this project we have
taken a distinct approach that takes advantage of this apparent
difficulty. Since all TNO detections will be trailed to some degree,
analyzing a single set of detections (the centroid of the trails) is
not the optimal method.  This trailing motivates the overall strategy
of our survey.  We explore the range of orbital parameters consistent
with the TNO region and keep those that produce significantly
different trails, keeping them as test trails. Then, the search for
sources in each image is optimized to select objects that show the
particular test trail. Sources are then correlated in the same way
ground-based observations are, and considering motion from image to
image that is consistent with the test trail considered.

\subsection{Detection using Optimized Kernel Search}
Searching for all possible orbits in the trans-neptunian space
requires an algorithm capable of sampling the complete set of
observational features that real TNOs could exhibit. This usually
translates into a set of possible rates in R.A.  and Dec. that are
surveyed with a rate resolution finer than that set by ${\rm FWHM} /
\Delta t$, where $\Delta t$ is the time span of the observations. For
HST data this is a bit more complex than ground based
observations. The extra parallax due to the motion of HST around the
Earth is $\sim0\farcs4$, significantly larger than the astrometric
uncertainty ($0\farcs05$) implying some structure in a single
detection can be identified even for TNOs.

This motivated us to use an optimized kernel search. Instead of taking
the point-source catalog for every image and searching those for
position correlations consistent with any orbit we consider a set of
orbits and search the images for detections consistent with them.

Each search is performed on the convolution of the original image and
a kernel designed to match the signature of an object with the orbit
being surveyed. This has the advantage of lowering the number of
artifacts while increasing our sensitivity to moving objects. The use
of kernels does not affect the photometry as it is only used for
detecting sources. The kernels were computed on the fly in the same
way a moving object is implanted, fitting a 2nd-degree polynomial to
its motion and implanting a set of 1-pixel separated PSFs on that
track. The detection was performed using SExtractor
\citep{Bertin.1996} that has built in the use of kernels.  The number
of orbits considered depends on the kernels; if two orbits produce
kernels that differ by less than a pixel in all images then only one
is used.

For every orbit considered a catalog is generated and fed into our
search algorithm. This takes the same set of orbital parameters used
to create the kernel and constructs a shift matrix $\delta_{ij}$, that
indicates by how much would an object with the current orbit in image
$i$ move from image $j$. Every detection in all images is at
considered as a possible moving object detection, and all other
detections are tested for a possible link with it. Detections are
linked by proximity to the putative new position using a threshold
equivalent to the astrometric precision.

Only links of three or more detections are considered viable moving
objects. We then filter solutions so that any detection belongs only
to one possible moving object, using the astrometric error of the
orbital solution with respect to the detections as the parameter to
rank these links. The result is a list of candidate objects, each
characterized by a set of detections.

\subsection{Visual Examination}\label{sec:vis}
At this stage the list of candidate objects is presented to a human
operator to distinguish moving objects from spurious detections. Up to
this point the pipeline is fully automatic with no step in the
reduction process requiring the input of an operator. The list
produced is the pipeline's best guess for which detections appear to
be related by a plausible orbit. However, no matter how efficient the
processing might be, the whole pipeline relies on the positional
information derived by Sextractor. It is nearly impossible to avoid
chance alignment of spurious detections or poorly subtracted cosmic
rays, for example, and to program an automatic selection algorithm
that could flag these events would be even harder.

The human brain is incredibly good at finding patterns. We make use of
this fact by presenting the detections as a pattern recognition
problem to a human operator. Each candidate is represented by an
animated postage stamp of the area around its location in the original
image and the one that was CR-removed, both with the detections
clearly marked. Both images are embedded in an webpage that gives the
option of flagging the object as moving object or as an
artifact. Information about the detections are also made available to
the observer. It usually takes $\sim3\min$ to a trained operator to
flag all objects in a field as moving or artifact.

On average the operator is presented with $\sim100$ objects per
\pointing~and nearly half of the detections that go through the human
filter were recognized as artifacts. These usually correspond to:
chance alignment of cosmic-rays (readily recognized due to their poor
fit and for appearing much brighter than reported), extended objects
elongated in the direction of the ecliptic (galaxies, saturated stars'
wings), etc.

It is only now that the list of selected objects is compared to that
of implanted ones. Those that are related to a synthetic object are
used to characterize the detection efficiency of our method, and those
that are ``real'' moving objects are flagged for constructing the
luminosity function. In all \pointings~considered in this project we
recovered over 5,000 fake objects, many times more than the 14 real
objects discovered. The fact that real objects (apart from a binary)
were indistinguishable from the implanted objects is a sign that the
search is well described by our control population.

\section{Detection Efficiency}\label{sec:eff}
After the list of implanted objects is revealed and we correlate it
with that of the objects found, the next step is computing the
efficiency function. The likelihood of obtaining a particular set of
objects from the control population depends on the efficiency function
$\eta(R)$. This likelihood function ${\mathcal L_\eta}$ has the form:
\begin{eqnarray}
{\mathcal L_\eta} & = & \prod_{i=1}^{N^+} \eta(R_i) \times
\prod_{j=1}^{N^-} \left[ 1-\eta(R_j) \right] \\
\eta (R) & = & \frac{A}{2} ~~ {\rm erfc} ~\left[ \frac{R-R_{50}}{2~w}
  \right] ~~,
\end{eqnarray}
the probability of finding a set of objects ($1,\ldots,N^+$) and of
not finding the complement ($1,\ldots,N^-$). The parameters in $\eta$
are the maximum efficiency ($A$), the magnitude at which the detection
probability equals half that of the maximum ($R_{50}$) and the width
of the decline in probability ($w$). We search for those values that
maximize ${\mathcal L}_\eta$.

In general each \pointing~may be considered as an independent survey
with its own detection efficiency and an area equal to ACS/WFC's field of
view. However, this is only true for uncorrelated observations, where
there is no chance of finding the same object in distinct
\pointings. Since we are using archival data, there are many
consecutive observations of the same field that were included in our
survey, where the probability of ``discovering'' an object twice is
non-negligible. If this effect is not considered appropriately we will
over estimate the area surveyed and consequently underestimate the TNO
luminosity function.

This problem may be solved if we know how many of the real TNOs would
move from \pointing~to \pointing. This requires that we have an
accurate model for the orbital distribution of the TNO population,
since different distributions will yield different levels of
``contamination''. The main variable determining how much an object
appears to move is its distance to the observer. We used a
distribution that resembles the heliocentric distribution in
\cite{Fuentes.2008}. The density of trial objects will determine the
statistical significance of the overlapping areas between \pointings.

We could use a single population with an accurate orbital distribution
and a high density for the entire area of the sky. However, as the
field of view of all our targets is negligible compared to the area
from which they were selected a density of $10$ per \pointing~would
yield $\sim10^8$ objects for which orbits and ephemeris would have to
be computed. Additionally, a different control would be necessary to
test the detection efficiency of uncommon, but physically plausible
orbits.

Instead we chose to separate the problem into detection efficiency and
effective area. The detection efficiency is well sampled for every
\pointing, as described above. In order to account for the area that
was observed by more than one \pointing~we find all intersections
between related \pointings. The area that intersects two
\pointings~corresponds to the fraction of the area in a field where a
TNO could have been detected twice. Since we are dealing with moving
objects we need to take into account the orbit distribution of the
real TNO population. We use a swarm of fake bodies in each
\pointing~to estimate the overlap. We first identify plausibly
correlated \pointings~by their observing time and location, 64 such
sets were found. We then created a large population of mock orbits
(1,000) with similar characteristics to the real TNO population in
each one of those \pointings~and computed how many fell in the field
of view of each other. The result becomes a bit more complicated when
we consider that an object could be in 3 or more of those \pointings,
each with its own efficiency function. In our survey we had a maximum
of 5 \pointings~that were correlated and could identify and precisely
account for all intersection areas that were surveyed more than
once. The effective area ($\Omega_\eff$) in \Fig{fig:cumfunc} has over
2,000 parameters and was computed as shown in \Eq{eq:eff}.
\begin{eqnarray}\label{eq:eff}
\Omega_\eff & = & \sum_{i=1}^{N} \sum_{S\in {\mathcal P_i}} \Omega_{S} \eta_{S} \\
\eta_{S}    & = & 1 - \prod_{o\in S} \left( 1 - \eta_{o} \right) ~~,
\end{eqnarray}
where the sum indexed $i$ is carried out over all $N$ sets of related
\pointings. If a set has $n_i$ \pointings, then there are $2^{n_i}$
possible combinations of overlapping fields or subsets $S$ in its
power set ${\mathcal P}_i$. Each one of those subsets represent an
area $\Omega_{S}$ that was surveyed with a detection efficiency
$\eta_{S}$. The detection efficiency of the subset is the probability
of being detected in any of the \pointings~in it. The computation of
$\Omega_{S}$ is provided by the fake bodies. It is the same fraction
of a field's area as the fraction of objects created in any of the
\pointings~in $S$ that end up in all \pointings~in $S$.

Given the large number of fields and filters considered, and despite
the many considerations the shape of the effective area does not vary
too sharply compared to the total area and can be approximated as a
function with only four parameters:
\begin{equation}\label{eq:effapprox}
\Omega_\eff(R) \approx \frac{A}{4} ~{\rm erfc} ~ \left[
  \frac{R-R_{25}}{2~w_1} \right]
~{\rm erfc} ~\left[ \frac{R-R_{25}}{2~w_2}\right] ~~,
\end{equation}
where the maximum effective area $A = 0.28 \pm 0.01 \deg^2$, the
magnitude at which the detection efficiency is 25\% of its maximum
$R_{25} = 26.5 \pm 0.1$, and the width of the decline in efficiency is
parameterized as $w_1 = 0.78 \pm 0.3$ and $w_2 = 0.31 \pm 0.3$. In
similar surveys it is conventional to define the magnitude at which
$\Omega_\eff(R)$ is half its maximum, which for our survey is
$R_{50}=26.14$. The maximum effective area $0.28\deg^2$ is comparable
to the total area surveyed $0.45\deg^2$. These differ due to the many
fields that effectively sampled the same objects. The effective area
is shown in the top panel of \fig{fig:cumfunc}.

After the real objects are recognized and some members of the control
population are identified among the detected objects we analyze the
photometry and astrometry of each one of those detections. We
construct the efficiency function and luminosity function. The orbital
constraint on every object is also investigated to understand the
uncertainties and possible degeneracies imposed by the data.

\section{Analysis}\label{sec:ana}

\subsection{Photometry}\label{sec:phot}
After sources are detected SExtractor \citep{Bertin.1996} is used to
obtain photometry. As was discussed in \S\ref{sec:dat} TNOs will shift
their position during the exposure and their shape will be elongated
in the direction of motion. We selected the {\tt AUTO} flux
measurement since it is the most appropriate for extended objects. It
uses an elliptical aperture which is computed for every detected
source. Instrumental magnitudes are provided for each discovered
object in \tab{tab:obj}.

For the range of magnitudes that is relevant for this study the
photometric uncertainty is dominated by the noise in the background
illumination. This uncertainty is also computed by SExtractor and is
in good agreement with the deviation between different images and with
the error for implanted objects. The photometric accuracy depends
mainly on the background brightness and the filter used.

The suite of filters that we considered for this project is presented
in \tab{tab:filters}. Transformations between ACS/WFC magnitudes and
UBVRI standard magnitudes were computed based on \citep{Sirianni.2005,
Jordi.2006} and considered typical colors for TNOs based on
\cite{Doressoundiram.2008} (V-R $=$ 0.6; R-I $=$ 0.6; B-R $=$ 1.6)

\begin{deluxetable}{llcc}
\tabletypesize{\scriptsize}
\tablecaption{\sc Photometric conversion}
\tablewidth{0pt}
\tablehead{  \colhead{$Filter$} &  \colhead{$Description$} &  \colhead{$zeropoint$}  &  \colhead{$R - Filter$}}
\startdata
F435W & Johnson B & $25.17$ & $-1.02$ \\
F475W & SDSS g'   & $25.77$ & $-0.54$ \\
F555W & Johnson V & $25.69$ & $-0.66$ \\
F606W & Broad V   & $26.67$ & $-0.61$ \\
F625W & SDSS r'   & $26.23$ & $-1.03$ \\
F775W & SDSS i'   & $26.42$ & $-0.65$ \\
F814W & Broad I   & $26.80$ & $-0.69$ \\
F850LP & SDSS z'  & $25.95$ & $+0.32$
\enddata
\tablecomments{HST filter name, equivalent standard name, and their
respective zeropoint. The transformation to $R$ assumes TNO colors
(V-R $=$ 0.6; R-I $=$ 0.6; B-R $=$ 1.6.)  }
\label{tab:filters}
\end{deluxetable}

\subsection{Orbital Information}
Though we use the position of the objects in each image and its trail
over a single exposure to find orbits that are consistent with an
objects' motion, we obtain a tighter constraint if we simulate the
images themselves. We use the stability of HST's PSF, and its angular
resolution to find the set of orbital parameters that are consistent
with the data and in this way provide accurate uncertainty estimates
for them.

Our ability to constrain the range of orbital parameters for a given
object is greatly improved by the motion of the telescope during a
\pointing. The extra parallax and precise astrometry provided by HST
allow us to better disentangle the parallactic and proper motion of an
object during the exposure. The motion of HST will produce a parallax
for any motion perpendicular to the ecliptic that will be evident as a
curved path in the image. For fields at low ecliptic latitude the
component along the ecliptic is largest and changes with time, as the
target ``rises'', ``transits'', and ``sets'' with respect to Earth
throughout the \pointing. On images with equal exposure time this will
manifest as a set of streaks with different lengths.

We run a Markov Chain Monte Carlo (MCMC) simulation where the function
to minimize is the residual on the objects' image. To compute the
$\chi^2$ we consider a rectangular region around each detection that
depends on the shape of the trail and the uncertainties in the
data. We parameterize this function on variables related to the
observed motion on the sky. We constrain the number of parameters we
fit for to those that affect the orbit of the objects, in order to
speed the convergence of the Markov Chain, hence fluxes are taken from
SExtractor. We use the best ``test orbit'' from the automatic process
as the starting point to define the section of the images where the
model and residuals will be computed. This allows the inclusion of all
images in the \pointing, regardless of whether SExtractor found the
object in every image or of any error in the position of the
detections.

The utility and success of MCMC is greatly increased if we are able to
find a transformation to a set of orthogonal variables, where the
effect on the target likelihood produced by a small change in one
dimension is decoupled from changes in others. Using Keplerian
elements is not the most appropriate choice of variables since a small
change in one element would affect others if the position of the
object at a given time is to remain constant, making our method very
inefficient. For this part of the analysis we considered the
parameterization and routines developed by
\cite{Bernstein.2000}. These consider a cartesian coordinate system
centered on the observer that points toward the center of the first
image. Though its variables are fairly independent when describing the
parallactic motion of a TNO, we took into consideration a few
modifications to the parameters. These changes of variable were chosen
to ensure a smooth transition between different areas of the parameter
space that yield similar trajectories. Since a change in distance also
changes the travel time we were forced to include a shift in the
object's relative position in every step so that the Markov Chain
would not get stuck updating all other parameters every time the
distance changed.

\begin{figure*}[Ht]
  \epsscale{1.01} \plotone{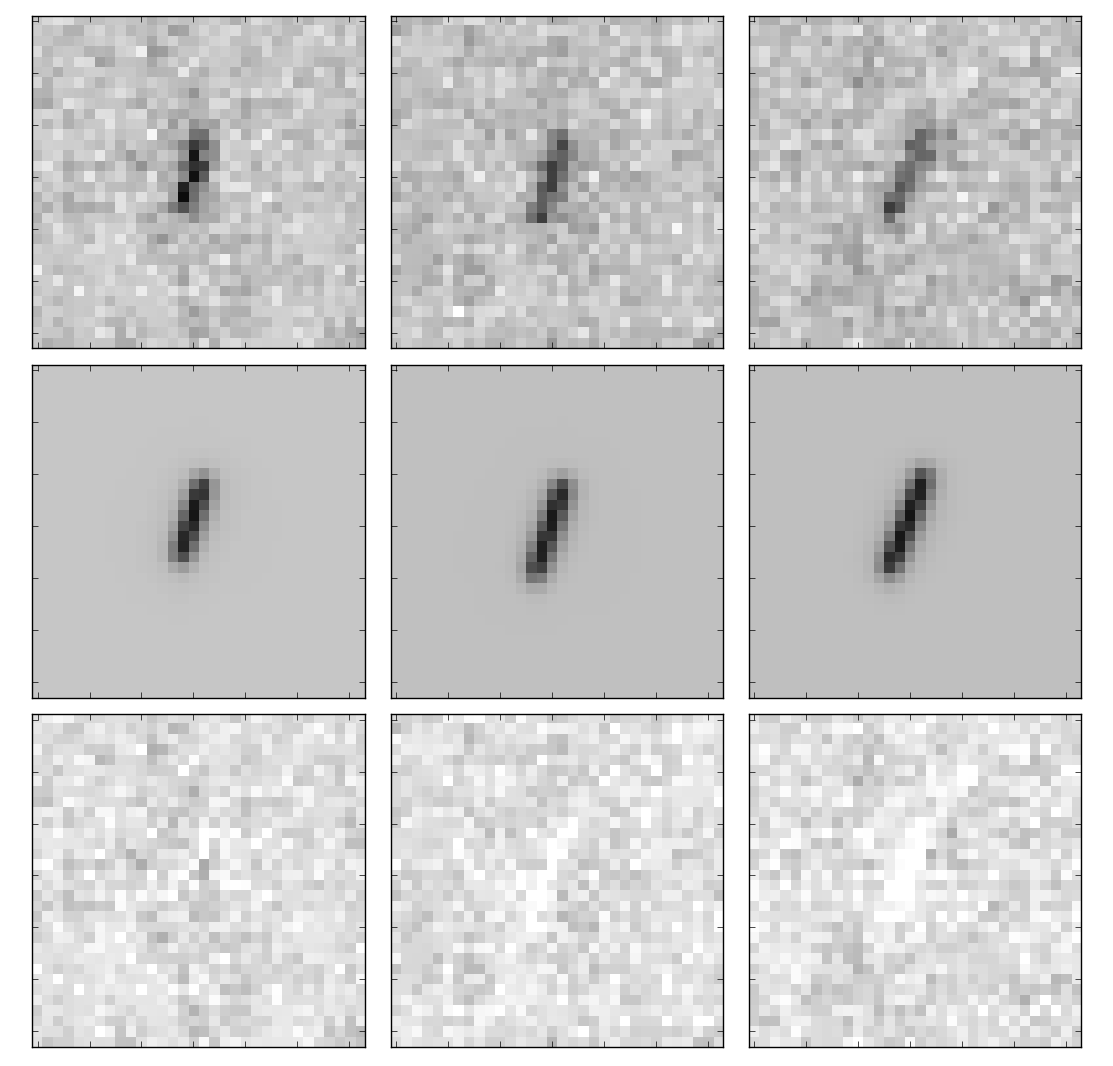} 
  \caption{\label{fig:model} Images around the location of the found
    object $hst11$ in each of the three images in the \pointing~where
    it was found. Each row shows the data after cosmic ray processing,
    the model and the residuals for an acceptable trial sampled during
    the MCMC minimization of the residuals. Note that we did not fit
    for the fluxes but took them from SExtractor. }
\end{figure*}

The only constraint we imposed on trial orbits was that they were
bound and that the velocity along the line of sight was zero, a good
approximation given the short arc and the large distance to these
objects. In \fig{fig:model} we show the postage stamps around one of
the objects $hst11$ for an iteration in the Markov chain.

\subsection{Binarity}
\begin{figure*}[Ht]
  \epsscale{1.01} \plotone{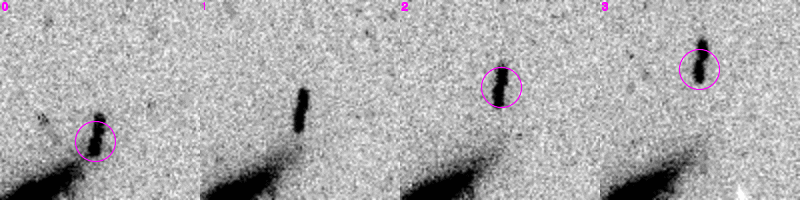} 
  \caption{\label{fig:binary} A postage stamp sequence of cosmic ray
  corrected images around the position of $hst5$. No
  distortion correction has been applied to these images. The
  detections that were linked by the search algorithm are shown as
  magenta circles. The component closer to the background galaxy has a
  F814W magnitude of $23.6\pm0.3$ and the other one $23.7\pm0.3$. The
  separation is $\delta \alpha=0\farcs53 \pm 0.01$, which at a distance
  of $42.9\pm0.6 \au$ gives a lower limit to their physical separation
  $a > 165,000 \pm 2,000 \km$.  }
\end{figure*}

In our sample of 14, there is only one object that is readily
recognized as binary (See \Fig{fig:binary}). The separation between
the components is $\delta \alpha=0\farcs53 \pm 0.05$ and their
magnitudes are: $23.6\pm0.3$ and $23.7\pm0.3$ respectively, making
this a very likely equal-mass binary.

The characteristics of this binary are quite common among the binary
TNO population. Its absolute solar system magnitude $H\sim6.8$ (a
proxy for size) and inclination ($i\sim3.5$) place it among many other
binaries in \cite[Figure 2]{Noll.2008}.

\Fig{fig:binary} shows an obvious binary, however limits on the binary
fraction are difficult to obtain, given that we did not calibrate our
search for binary detection. No binary control population was
implanted and for this reason we are unaware of our efficiency at
detecting them as a function of separation, brightness ratio or
orbit. Nevertheless, having one detection we can only place a $7\%$
lower limit on the fraction of wide, equal brightness binaries among
the faint TNO population.

\subsection{Size distribution}\label{sub:siz}
\begin{figure*}[Ht]
  \epsscale{1.01} \plotone{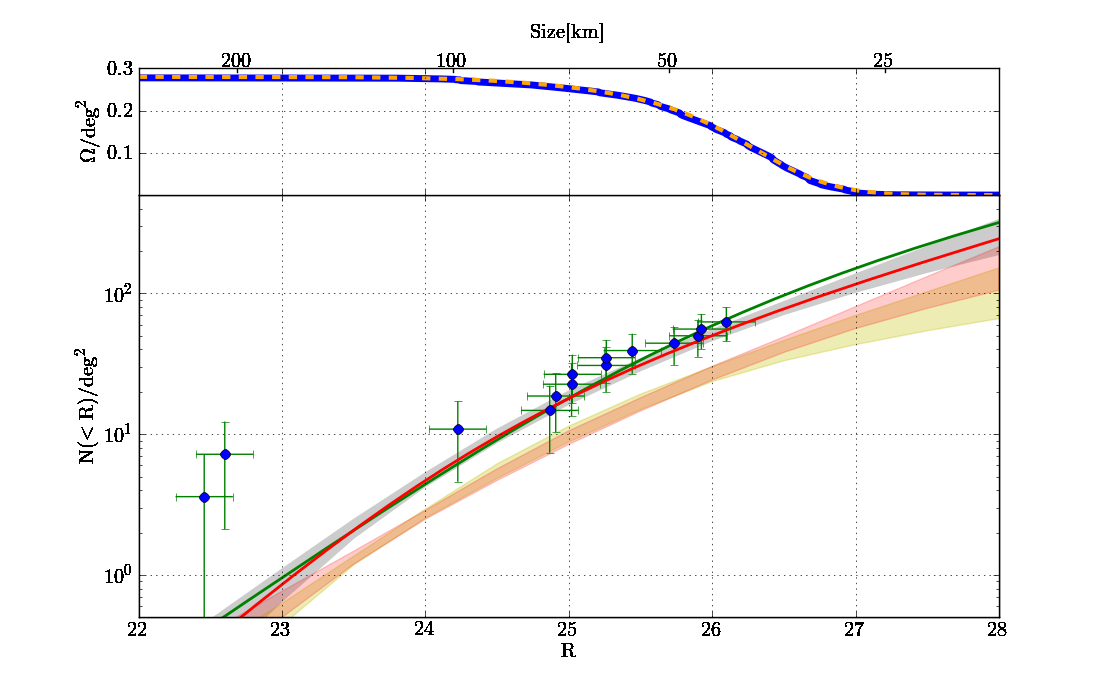} 
  \caption{\label{fig:cumfunc} The top panel shows the effective area
    surveyed in this paper as a function of $R$ magnitude in
    blue. While that function depends on the efficiency function and
    shared area of every \pointing~in this survey we can simplify the
    $\sim2,000$ parameters (See \eq{eq:eff}) into the 4-parameters of
    the function plot in dashed orange (See \eq{eq:effapprox}). The
    lower panel shows the luminosity function of objects found in this
    survey, normalized by the effective area at each magnitude. The
    best model in \cite{Fuentes.2009} is overplotted in green, while
    the best model for all surveys, including this one, is shown in
    red. The gray area represents the area enclosed by the $1-\sigma$
    confidence region for all surveys. The lower set of shaded areas
    represent the $1-\sigma$ confidence limits for the cumulative
    function of the hot (red) and cold (yellow) population. }
\vspace{0.1in}
\end{figure*}

Once the distance and magnitude are measured we can compute the size
of each body, assuming a value for the albedo. If we further assume
all objects are roughly at the same distance, the luminosity function
can be written as a function of size, as shown in \fig{fig:cumfunc}
where we assume the distance $d=42\au$ and the albedo $p=0.07$
\citep{Stansberry.2008}.

Introduced in \cite{Bernstein.2004}, the Double Power Law (DPL) is a
handy functional form for the density of objects as a function of $R$
magnitude that considers a break in the size distribution. The
parameters are $\sigma_{23}$ or the surface density of objects with
$R=23$, $\alpha_1$ and $\alpha_2$ or the slopes of the power law
behavior of the luminosity function for the brightest and smallest
objects, and $R_{eq}$ is the magnitude where the behavior changes from
that of small to that of large sizes.

\begin{figure*}[Ht]
  \epsscale{1.01} \plotone{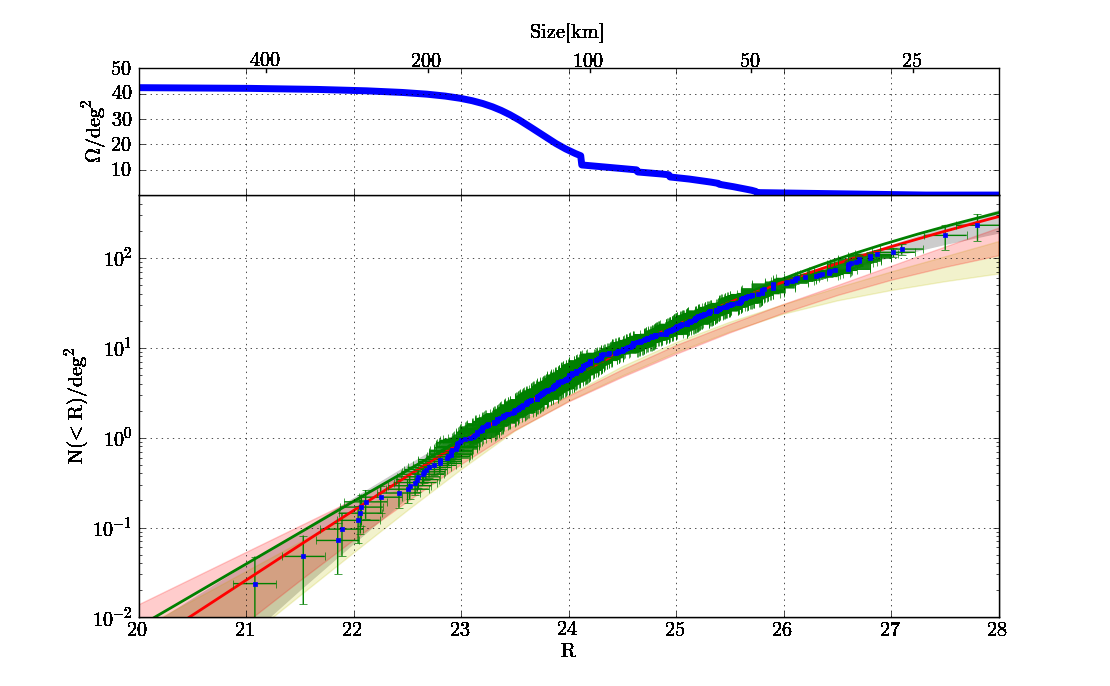} 
  \caption{\label{fig:cumfunc_all} The top panel shows in blue the
    effective area for all surveys considered in this paper as a
    function of $R$ magnitude, normalized by the effective area at
    each magnitude. The lower panel shows the luminosity function of
    TNOs in all surveys considered, normalized by the effective area
    at each magnitude. The best model in \cite{Fuentes.2009} is
    overplotted in green, while the best model for all surveys,
    including this one, is shown in red.  The gray area corresponds to
    the $1-\sigma$ confidence region given for all objects. The same
    confidence regions are given for hot and cold objects, in red and
    yellow respectively.}
\vspace{0.2in}
\end{figure*}

The previous best fit to the cumulative size distribution, that
combined the results of all surveys listed in \cite{Fuentes.2009} is
shown in \fig{fig:cumfunc} in red. We also considered our 14 objects
together with the many surveys that provide detailed information about
their calibration (See Table 2 in \cite{Fuentes.2008} and
\cite{Fuentes.2009, Fraser.2009}.) The total area surveyed or
effective area in all those surveys is plotted in the top panel of
\fig{fig:cumfunc_all}. We consider only objects that were discovered
at magnitudes brighter than the magnitude at which their respective
surveys' detection efficiency fell below 15\% of the maximum
efficiency.

Using the orbital information provided in those surveys, we define the
hot and cold populations as those with inclinations larger and smaller
than $5^\circ$. A caveat about some of the surveys that provide
inclination and distance information with a $\sim24$-hr arc at
opposition is that they can only compute a rate of motion on the sky.
\cite{Fuentes.2009} included only the rate of motion for every object,
from which we computed the distance and inclination.

\begin{figure*}[Ht]
  \epsscale{1.01} \plotone{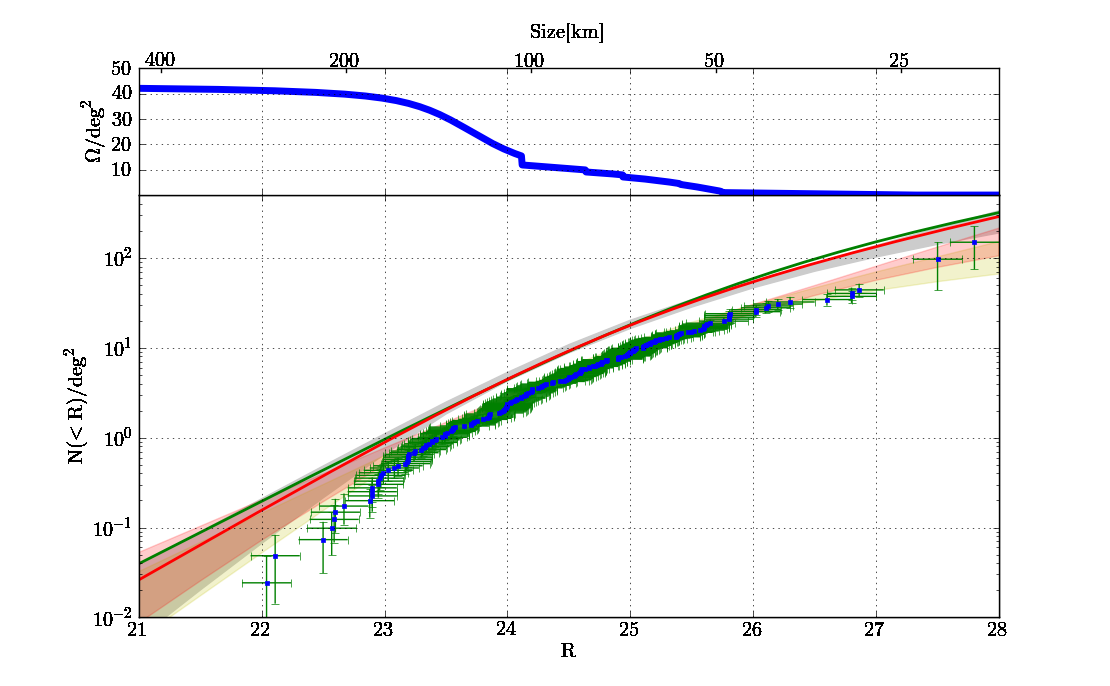} 
  \caption{\label{fig:cumfunc_cold} Same as \fig{fig:cumfunc_all} but
  only for objects deemed dynamically cold, $i\leq5^\circ$}
\end{figure*}

\begin{figure*}[Ht]
  \epsscale{1.01} \plotone{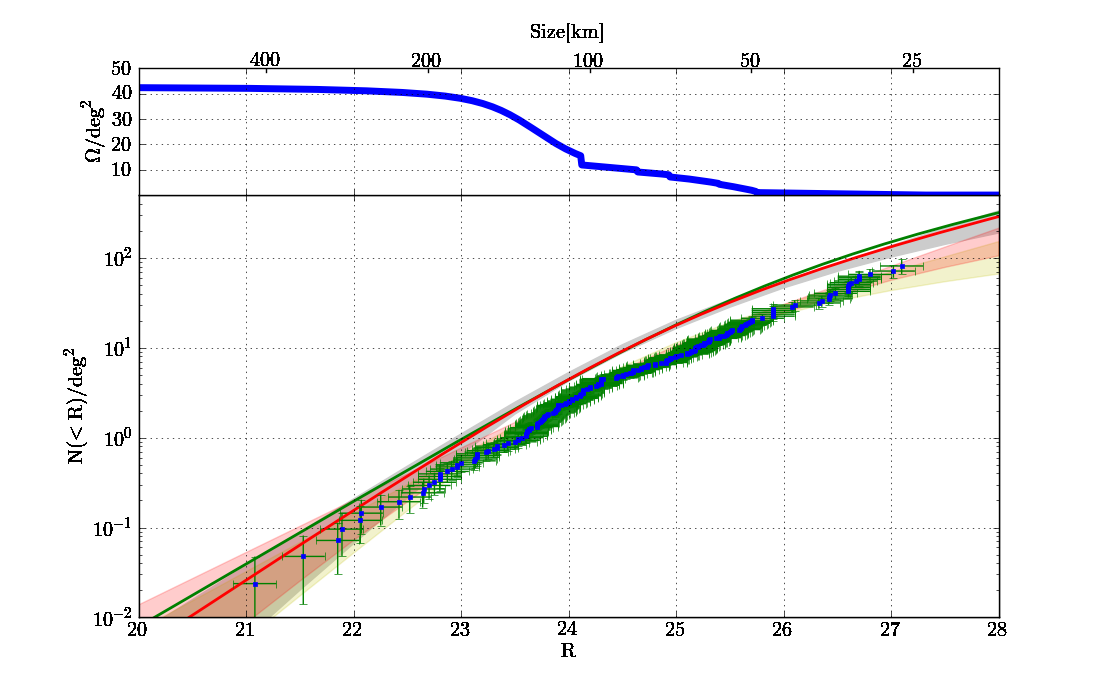} 
  \caption{\label{fig:cumfunc_hot} Same as \fig{fig:cumfunc_all} but
  only for objects deemed dynamically hot, $i>5^\circ$}
\end{figure*}

\begin{figure*}[Ht]
  \epsscale{0.6} \plotone{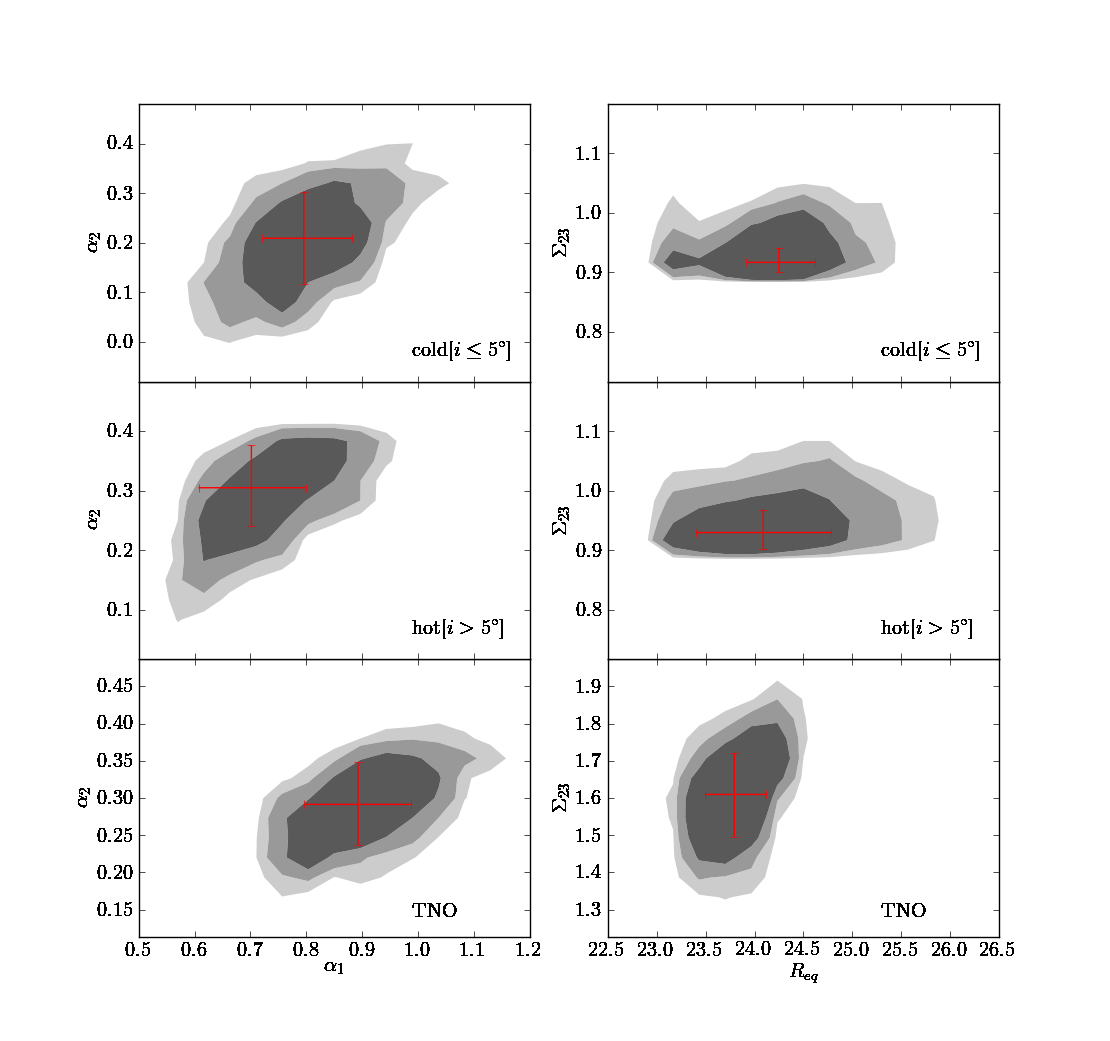} 
  \caption{\label{fig:lumfunc_prob_all} Probability density for the
    surface number density $\sigma(R)$. The parameters are those in
    the double power law model. In each panel the painted areas
    represent the 3, 2 and 1-$\sigma$ confidence regions. The panels
    on the left show the probability distribution as a function of the
    power law exponents for bright and faint objects ($\alpha_1$,
    $\alpha_2$). Panels on the right axis show the likelihood of the
    brightness at which the luminosity function changes slope,
    ($R_{eq}$) and the density of objects at $R=23$
    ($\Sigma_{23}$). The red crosses show the $1-\sigma$ confidence
    region for each parameter when the probability density has been
    marginalized over all other variables. The details of this
    likelihood analysis can be found in \cite{Fuentes.2008} and
    references therein.  The bottom panel shows the results for all
    objects in all surveys considered, the most likely value for the
    parameters is: $\alpha_1 = 0.89\pm0.10$, $\alpha_2 = 0.29\pm0.06$,
    $\Sigma_{23} = 1.61\pm0.11$, $R_{eq} = 23.8\pm0.3$. The middle
    panel shows only objects considered as hot or excited, selected
    for having $i>5^\circ$, the best parameters are: $\alpha_1 =
    0.70\pm0.10$, $\alpha_2 = 0.30\pm0.07$, $\Sigma_{23} =
    0.93\pm0.03$, $R_{eq} = 24.1\pm0.7$. The top panel corresponds to
    cold objects ($i\leq5^\circ$), where the most likely solution is
    $\alpha_1 = 0.80\pm0.08$, $\alpha_2 = 0.21\pm0.09$, $\Sigma_{23} =
    0.92\pm0.02$, $R_{eq} = 24.2\pm0.4$. }
\end{figure*}

We analyzed the likelihood function for all these observations given
the effective surveyed area following the MCMC analysis described in
\cite{Fuentes.2008}. The likelihood function is plot against two of
the DPL variables in each of the panels in \fig{fig:lumfunc_prob_all}
for the hot, cold and all objects (top, middle and lower panel,
respectively). The constraints on the DPL parameters, computed on the
likelihood of each parameter marginalized over all others is:
$\alpha_1 = 0.89\pm0.10$, $\alpha_2 = 0.29\pm0.06$, $\Sigma_{23} =
1.61\pm0.11$, $R_{eq} = 23.8\pm0.3$ for all objects considered,
$\alpha_1 = 0.70\pm0.10$, $\alpha_2 = 0.30\pm0.07$, $\Sigma_{23} =
0.93\pm0.03$, $R_{eq} = 24.1\pm0.7$ for hot objects and $\alpha_1 =
0.80\pm0.08$, $\alpha_2 = 0.21\pm0.09$, $\Sigma_{23} = 0.92\pm0.02$,
$R_{eq} = 24.2\pm0.4$ for cold ones.

\section{Discussion}\label{sec:dis}
Our search can be compared to the targeted use of HST by
\cite{Bernstein.2004} where 6 fields where imaged $\sim20$ times each
to find the faintest TNOs possible. Though that search was
significantly more sensitive to faint objects due to a careful
selection of the fields, the coaddition of signal and the use of a
wide-filter, it only focused on six independent fields. That group
found 3 objects, two of which remain the faintest TNOs ever
imaged. This work would have been able to detect the two brightest of
those objects in any \pointing~that satisfied our criteria. A more
targeted survey, with better selection of filters and \pointings~at
the stationary point, where objects won't trail as much, would be much
more efficient at finding TNOs than our archival search.

Of the total area imaged, $0.45\deg^2$ corresponding to the 150
\pointings~that were analyzed, the effective area for this survey is
only $0.3\deg^2$, as shown in the top panel of \fig{fig:cumfunc}. This
is due to background sources, cosmic ray confusion, and any other
feature that would completely or partially prevent us from detecting
an object in at least three images. The main cause for this reduced
survey area is, however, the existence of \pointings~of the same field
taken is succession, which effectively increases the chances of
detecting an object in that field but decreases the area of our survey
by re-observing a field, where the same objects are visible, many
times.

The luminosity function of the 14 objects discovered in this survey is
presented in the lower panel in \fig{fig:cumfunc}. The effective area
is also plotted in the top panel to put the significance of each
detection in context. The previous best fit to the TNO population is
plot in green, the best fit for all surveys (including this ones) is
also plotted in red. As we can see our survey follows precisely the
expectations derived from previous work. There are two bright objects
$R<23$ among the 14, which indicates a higher density than
expected. The statistical significance of this deviation is low.

We took most calibrated surveys for TNOs in the literature (See
\S\ref{sub:siz}), along with this one, to construct a effective survey
area and luminosity function for all these objects, shown in
\fig{fig:cumfunc_all}. There are over 400 TNOs included, which allows
a precise constraint on the luminosity function. The two faintest
objects beyond $R=27.5$ were discovered with HST
\citep{Bernstein.2004}, and we see that ground based surveys are
already sensitive to $R=27$ TNOs.

The exquisite astrometric precision in HST data enables us to measure
a TNO's distance and inclination with a few percent uncertainty, even
from an arc as short as 40 min. Although an object's motion depends on
the solar elongation at which the observations were taken, the extra
motion due to HST's orbit helps disentangle the objects' parallax from
its proper motion along the ecliptic. Though this is a sample of only
14 objects, the distances found are in good agreement with
\cite{Kavelaars.2008}. Dynamically cold objects are constrained to
~35<d<50 AU, while hot objects do not cluster at a particular distance
(See Table 2).

Using the inclination information in \tab{tab:obj} we separate objects
into the hot and cold dynamical classes, for this we use a simple
$i=5^\circ$ cutoff. Applying this filter to all surveys we compute the
luminosity function of hot and cold objects (shown in
\fig{fig:cumfunc_cold} and \fig{fig:cumfunc_hot} respectively.) We see
that the bright-end and faint-end slopes for the hot and cold
populations are similar. This is best seen in the MCMC posterior
probability for the luminosity function DPL parameters (See
\fig{fig:lumfunc_prob_all}.)  The hot and cold luminosity function
constraints on $\alpha_2$, the faint-end slope, are consistent with
each other. However, there is a significant deviation at the bright end
of the dynamical hot population between the best fit model and the
luminosity function which indicates that the size distribution of
bright hot objects is shallower than that of the cold fraction, the
largest and brightest objects tend to be dynamically excited,
consistent with the results of \cite{Levison.2001,
  Bernstein.2004}. Our lack of data to contrain the bright end of the
TNO luminosity function is explained as most of the objects in our
analysis come from well characterized pencil beam surveys where only a
few large objects are present. For the same reason, our constraints
are more significant for the faint end of the luminosity function.

This result is in contradiction with claims that the luminosity
function of hot and cold objects differs for small bodies
\citep{Bernstein.2004, Fuentes.2008}. However, there is an explanation
for this difference. For a long time the only TNOs fainter than
$R\sim26$ were those found by \cite{Bernstein.2004}, all of them
cold. This lack of faint hot objects allowed for extremely flat slopes
for smaller sizes. By including the deeper surveys of
\cite{Fuentes.2009} and \cite{Fraser.2009}, which have detected
several high inclination objects, the non-detection of $R>27$ hot
objects is less significant.

However, these deep ground based surveys rely on short arcs and must
constrain their orbits to be circular to compute a distance and
inclination. This increases the probability of contamination between
hot and cold objects. This problem is less severe with detections
obtained from HST. In our survey we see roughly equal number of cold
and hot objects, which is consistent with the ground based
inclinations being accurate.

As was discussed earlier, the size distribution is intimately related
to the luminosity function. If we disregard the distance estimate and
assume all objects are located at $42\au$ and have a 7\% albedo the
transformation is direct and corresponds to the top axis in
\fig{fig:cumfunc_all}. The break magnitude that marks the transition
between the bright and the faint slope luminosity function becomes
then a break in the size distribution. Such a break is expected from
the collisional evolution the TNO population has undergone since these
objects formed \citep{Kenyon.2004, Kenyon.2008, Pan.2005}. We find the
location of such break to be consistent for both hot and cold
populations $D_{eq}\sim100\km$. In the model of \cite{Pan.2005} such a
large size corresponds to the largest object that has been disrupted
in the age of the Solar System. We note that this results relies on an
assumed distance and albedo for all objects, something that we know is
inaccurate for distances. The albedo is also likely to be different
as there seems to be a correlation with size and color
\citep{Stansberry.2008}.

The results discussed in this section constrain the TNO size
distribution at low ecliptic latitudes. As we cover more of the sky,
sampling the TNO population away from the ecliptic we will measure
density of objects as a function of latitude, as well as the relative
proportion of hot vs. cold objects. We shall be able to include more
objects in our analysis and hence put better constraints on the
location of the break in the size distribution, for different
dynamical families.

\begin{figure*}[Ht]
  \epsscale{1.01} \plotone{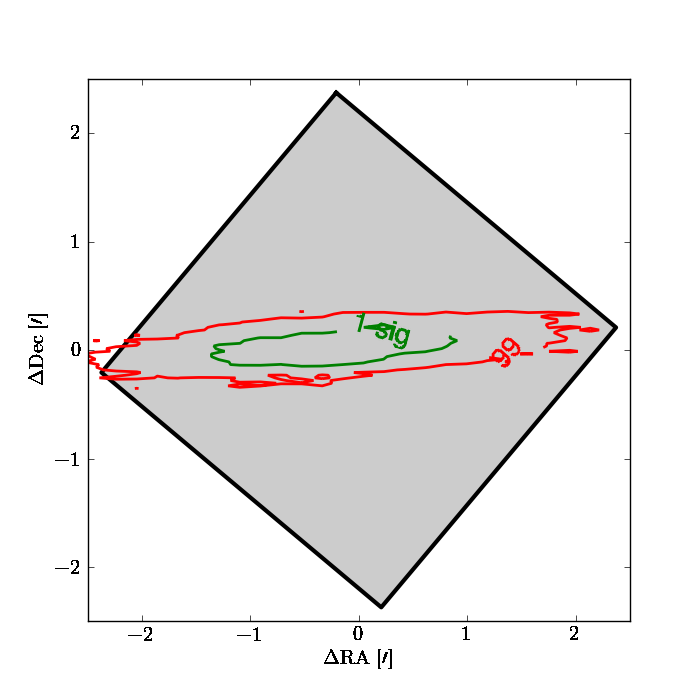} 
  \caption{\label{fig:unc} 1-sigma (green) and 99\% (red) uncertainty
    in the position of a particular object (120-deg solar elongation),
    30-days after discovery.  The field of view of ACS/WFC is
    overplotted as a black square. }
\end{figure*}

Among the objects found there is a wide-separation binary. Since we
did not consider the possibility of binaries in our control
population, we cannot directly extract the statistical significance of
our measurement. However, our result allows us to set a limit on the
binary fraction of $7_{-2}^{+13}\%$, in excellent agreement with the
current limits for different population of 5-20\% \cite{Noll.2008}.

\begin{figure*}[Ht]
  \epsscale{1.01} \plotone{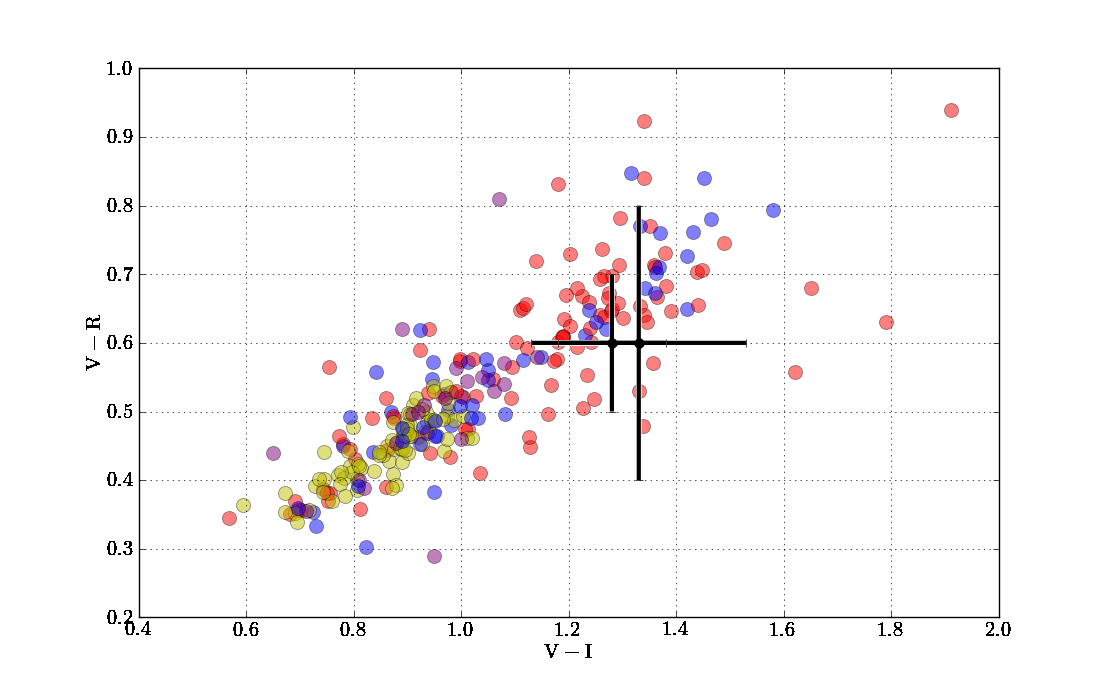} 
  \caption{\label{fig:colors} Colors for our two objects along with
    previously known TNOs in the mboss database \citep{Hainaut.2002}
    (red: classicals and plutinos;~ blue: scattered and centaurs;~
    yellow: trojans;~ purple: long and short period comets.)  The
    black points correspond to $hst13$ and $hst17$ respectively, where
    the V-R color for both is assumed to be 0.6 and the uncertainties
    are the same as those in V-I.
 }
\end{figure*}

For the couple of objects with two different filter observations we
put them in context of the \cite{Hainaut.2002} database, as shown in
\fig{fig:colors}. The two objects seem to fall right in what is
expected for classicals, which is consistent with their inclination
estimates: $i\sim$ ${\displaystyle 10^\circ}$ and ${\displaystyle
  5^\circ}$. For at least these two objects the assumption V-R=0.6 is
justified.

The objects presented in this paper typically cannot be followed
up. The short observation arc, and the long time elapsed since the
date the observations were taken imply an uncertainty in the position
too large to recover them in current observations.  However, the
orbital estimates are accurate enough to grant an uncertainty ellipse
that fits within an ACS/WFC fields within a month after the discovery
observation (See \Fig{fig:unc}.)  This opens the possibility of a
slightly different observing strategy, where data are processed as
they become available, and follow-up observations can be scheduled
quickly.

\section{Conclusions}\label{sec:con}
We have successfully completed a search for TNOs within 5 degrees of
the ecliptic using archival data taken with the ACS/WFC camera aboard
HST. The data span 6 years. Of the 150 \pointings~analyzed 14 objects
were found, yielding roughly 1 object per 10 \pointings. This suggests
that there are possibly hundreds of new TNOs with exquisite astrometry
and photometry still hidden in the ACS/WFC archive at higher ecliptic
latitudes. We have proven our ability to detect and characterize these
even with data intended for completely different purposes, where most
of the filters and strategies used for these observations are
sub-optimal for detecting TNOs.

Given the excellent astrometric precision of the images, it is
possible to use observations taken in a single \pointing~of HST to
predict the position of a TNO a month later and have the uncertainty
ellipse fit within the field of view of ACS. This, coupled to the fast
turnover of data from our pipeline may yield many viable candidates
for follow-up observations. A detection a month later would allow us
to collect a significant set of small TNOs with accurate orbits,
opening the possibility for detailed observations with present and
future instruments like JWST.

Binaries are only detected as such if the separation of the components
on the plane of the sky can be resolved. The trailing of the binary
imposes further constraints to the fraction of time a given binary
would be recognized, which is specially problematic for HST data where
trailing of TNOs is more prominent. Previous searches for binaries
using HST have surveyed known TNOs and tracked the telescope to
counteract their motion. The one relatively faint ($R\sim23$) binary
discovered in this project illustrates the successful detection of a
trailed binary.  This opens the possibility of constraining the rate
of binaries as a function of size from the same survey in which the
TNOs are discovered.

The recognition of a low and high inclination population among TNOs
\citep{Brown.2001} has been interpreted as the existence of two
dynamically distinct set of objects. Those with higher inclinations
or excited (hot) and those with lower inclinations and not excited
(cold). \cite{Bernstein.2004} found that these two, defined by their
inclination had different size distributions. There is also evidence
that they have different colors \cite{Doressoundiram.2008}.

The cold population's size distribution is steeper than that of the
hot population for objects larger than $\sim100\km$
\cite{Levison.2001}. In this paper we show that the location of the
break in the size distribution and its slope for objects fainter than
it for cold and hot objects are consistent with each other. This is
compatible with the theory of collisional evolution, where a
population with a given steep power law size distribution gets
collisionally grinded as time progresses \citep{Pan.2005, Kenyon.2004,
  Kenyon.2008}. The slope of the size distribution of small objects in
those models is constant since it is given by the steady state of
collisions. The location of the break is also consistent for both
populations, but $D_{eq}\sim100\km$ is larger than what theories
expect.

This difference could be better understood if any albedo dependence
with size is first investigated for smaller objects. At present we can
only extrapolate the apparent correlations between size and color, and
inclination and albedo from measurements performed only on the
largest, brightest TNOs. By chance, multifilter observations of two
TNOs were obtained, yielding V-I colors for two objects. We expect to
find more of these serendipitous color observations for other faint
TNOs in intensively surveyed fields away from the ecliptic.

The advent of the Wide Field Camera 3 (WFC3) installed in May 2009
opens the possibility for extending this work to the Near-IR and to a
larger fraction of the data collected by HST. The prospects of such
observations would allow extending surface studies to small objects.

As we continue the analysis of HST archival data to higher ecliptic
latitudes we will start sampling an area of the sky that has only been
surveyed for brighter ($R\lesssim21$ \citep{Trujillo.2003}. When the
whole archive is searched we shall take the depth and resolution of
pencil beam searches to the whole sky.

\acknowledgments Support for program 11778 was provided by NASA
through a grant from the Space Telescope Science Institute, which is
operated by the Association of Universities for Research in Astronomy,
Inc., under NASA contract NAS 5-26555.


\clearpage

\begin{landscape}
\begin{deluxetable}{llllccccccc}
\tabletypesize{\scriptsize}
\tablecaption{\sc Found Objects}
\tablewidth{0pt}
\tablehead{
  \colhead{${\rm Name}$} &
  \colhead{${\rm MJD}$} &
  \colhead{${\rm RA}$} &
  \colhead{${\rm Dec}$} &
  \colhead{${\rm Filter}$\tablenotemark{a}} &
  \colhead{$\overline{m}$\tablenotemark{a}} &
  \colhead{$R$} &
  \colhead{$H$} &
  \colhead{${\rm opp\_ang}$} &
  \colhead{$d$\tablenotemark{b}} &
  \colhead{$i$\tablenotemark{b}} \\
  \colhead{} &
  \colhead{} &
  \colhead{} &
  \colhead{} &
  \colhead{} &
  \colhead{} &
  \colhead{} &
  \colhead{} &
  \colhead{$[deg]$} &
  \colhead{$[AU]$} &
  \colhead{$[deg]$} \\
}
\startdata
hst4 & $53318.1470775$ & $11:06:02.741$ & $3:12:09.186$ & F775W & 25.7 & $25.0\pm0.1$ & $10.0$ & $60.5$ & $36.5\pm1.4$ & $2.9\pm0.4$ \\
hst5\tablenotemark{c} & $53585.640178$ & $22:15:01.360$ & $-13:59:46.710$ & F814W & 23.1 & $22.5\pm0.3$ & $6.8$& $160.9$ & $42.9\pm0.6$ & $3.5\pm0.2$ \\
hst6 & $53867.311035$ & $22:38:46.527$ & $-7:54:11.095$ & F814W & 26.6 & $25.9\pm0.3$ & $10.3$ & $73.1$ & $41.5\pm9.5$ & $6.9\pm4.5$ \\
hst7 & $53838.4348775$ & $12:29:27.110$ & $1:52:15.912$ & F850LP & 25.4 & $25.7\pm0.8$ & $9.0$ & $162.8$ & $53.8\pm1.9$ & $62.7\pm13.4$ \\
hst8 & $53964.9145558$ & $3:21:46.568$ & $16:52:34.607$ & F555W & 26.7 & $26.1\pm0.4$ & $8.3$ & $92.6$ & $69.1\pm33.3$ & $14.7\pm11.9$ \\
hst9 & $53956.060458$ & $3:21:46.559$ & $16:52:34.992$ & F814W & 26.1 & $25.4\pm0.2$ & $11.5$ & $84.2$ & $28.2\pm7.1$ & $18.4\pm16.4$ \\
hst10\tablenotemark{d} & $54034.98833$ & $3:21:46.716$ & $16:52:27.288$ & F625W & 25.9 & $24.9\pm0.3$ & $6.5$ & $160.9$ & $79.0\pm1.5$ & $28.6\pm21.4$ \\
                                                                                            & & & & & & & $6.2$ & & $85.5\pm3.3$ & $105.5\pm6.9$ \\
hst11 & $53786.0709287$ & $12:13:50.319$ & $2:48:43.607$ & F435W & 23.6 & $22.6\pm0.1$ & $7.8$ & $148.8$ & $34.7\pm2.5$ & $4.9\pm0.5$ \\
hst12 & $53989.5521546$ & $0:39:18.864$ & $0:53:11.789$ & F814W & 26.6 & $25.9\pm0.2$ & $5.3$ & $158.0$ & $131.9\pm4.5$ & $93.1\pm64.0$ \\
hst13\tablenotemark{d} & $53988.213526$ & $0:39:34.385$ & $0:51:40.193$ & F606W:F814W & 25.9:25.9 & $25.3\pm0.1$ & $9.5$ & $158.0$ & $44.0\pm1.9$ & $8.4\pm1.1$ \\
                                                                                                                        & & & & & & & $8.4$ & & $56.2\pm3.3$ & $172.9\pm2.0$ \\
hst14 & $54042.139279$ & $9:47:03.493$ & $10:06:25.987$ & F814W & 25.9 & $25.3\pm0.4$ & $10.4$ & $75.0$ & $35.1\pm1.0$ & $33.6\pm1.5$ \\
hst15 & $52764.837741$ & $21:55:05.532$ & $-9:22:01.181$ & F625W & 25.3 & $24.2\pm0.1$ & $7.5$ & $77.5$ & $53.6\pm1.7$ & $3.5\pm0.3$ \\
hst16\tablenotemark{d} & $52422.7726762$ & $13:54:15.318$ & $-12:33:23.226$ & F814W & 25.6 & $24.9\pm0.2$ & $8.6$ & $143.7$ & $48.5\pm1.4$ & $2.6\pm1.0$ \\
                                                                                               & & & & & & & $7.9$ & & $57.6\pm4.2$ & $173.3\pm6.3$ \\
hst17\tablenotemark{e} & $52941.37034$ & $22:14:53.188$ & $-14:00:40.282$ & F555W:F814W & 25.8:25.8 & $25.1\pm0.1$ & $9.2$ & $115.3$ & $44.6\pm2.5$ & $9.4\pm5.8$ \\
\enddata
\tablecomments{All objects found in this work are shown with their
  photometric and astrometric properties.  Positions given for the
  first detections. The barycentric distance $d$ and inclination $i$
  were estimated from a MCMC with a parameterization given by the {\it
    Orbfit} code \citep{Bernstein.2000}. Though some objects were
  discovered in the same field, the epoch of the observations is
  different. The Solar System magnitude $H=V+5\log{d~\Delta}$, a
  function of the $V$ magnitude $d$ and the distance to the observer
  $\Delta$, is computed assuming the phase angle is small and that the
  V-R color for all objects is 0.6. }
\tablenotetext{a}{ When multiple filters are used, a filter list and
  corresponding instrumental magnitudes are shown.}
\tablenotetext{b}{ When prograde and retrograde solutions are possible
  we report both peaks.}
\tablenotetext{c}{ This is the binary shown in \Fig{fig:binary}. }
\tablenotetext{d}{ The retrograde solution is presented in a second
  row; it is always at a larger distance from the observer than the
  prograde one. }
\tablenotetext{e}{ This object was found in two consecutive
  \pointings. We found the same solution fitting both orbits, though we
  do not include the results of the second \pointing~as one of the
  detections is right on the edge of the detector. }
\label{tab:obj}
\end{deluxetable}
\clearpage
\end{landscape}

\end{document}